\documentclass[aps,pra,showpacs,groupedaddress,twocolumn,amsmath,amssymb]{revtex4-1}

\usepackage[american]{babel}
\usepackage{graphicx}
\usepackage{times}
\usepackage{subfigure}
\usepackage{color}

\newcommand{\ket}[1]{\left|#1\right>}
\newcommand{\bra}[1]{\left<#1\right|}

\begin{document}

\title{Entangled Qubits in a non-Gaussian Quantum State}

\author{T. Kiesel, W. Vogel}

\affiliation{Arbeitsgruppe Quantenoptik, Institut f\"ur Physik, Universit\"at  Rostock, D-18051 Rostock,
Germany}

\author{B. Hage}
\affiliation{
ARC Centre of Excellence for Quantum-Atom Optics, Department of Quantum Science, The Australian National University, Canberra, Australian Capital Territory 0200, Australia
}

\author{R. Schnabel}

\affiliation{
Institut f\"{u}r Gravitationsphysik, Leibniz Universit\"{a}t Hannover
 and Max-Planck-Institut f\"{u}r Gravitationsphysik (Albert-Einstein-Institute),\\
Callinstrasse 38, 30167 Hannover, Germany
}

\begin{abstract}
We experimentally generate and tomographically characterize a mixed, genuinely non-Gaussian bipartite continuous-variable entangled state. By testing entanglement in 2$\times$2-dimensional two-qubit subspaces, entangled qubits are localized within the density matrix, which, firstly, proves the distillability of the state and, secondly, is useful to estimate the efficiency and test the applicability of distillation protocols. In our example, the entangled qubits are arranged in the density matrix in an asymmetric way, i.e.~entanglement is found between diverse qubits composed of different photon number states, although the entangled state is symmetric under exchanging the modes. 
\end{abstract}
\pacs{03.67.Mn, 03.65.Ud, 03.65.Wj}
\maketitle


\section{Introduction.} 

Since the early days of quantum mechanics, entanglement has been a topic of great interest~\cite{EPR, Naturw23-807}. Today, it is considered as the key resource for many applications, such as quantum computation, quantum communication and quantum cryptography, for recent reviews cf.~\cite{Horodecki-09,Guehne-09}. Last but not least, nowadays the relevance of entanglement is even discussed in the context of life sciences, cf. e.g.~\cite{Vedral}. Traditionally there is a distinct treatment of entanglement in two different regimes,
the discrete variable regime being based for instance on qubits, and the continuous variable (CV) regime having either Gaussian or non-Gaussian statistics, respectively. However, both regimes are connected to each other, since qubits can be constructed out of CV states~\cite{cvqubit}, and vice versa entangled CV states may be composed of entangled qubits. 

A quantum state is referred to as being entangled if it cannot be written as a statistical mixture of factorizable (uncorrelated) quantum states~\cite{Werner}. Although this definition is intuitively clear, it is hard to check in practice for general quantum states. For the case of CV states the Peres condition verifies an important class of entanglement by partial transposition (PT) of the density matrix~\cite{Peres}. The special case of Gaussian entangled states can be fully characterized by moments of second order~\cite{Duan,Simon}. Gaussian states are easy to prepare and it is easy to 
identify their entanglement. However, their application in quantum technology is limited. Non-Gaussianity of CV entangled states can be considered as a resource, which is of great interest for various applications in quantum information technologies.
A general reformulation of the Peres criterion yields a hierarchy of inequalities in terms of observable moments of arbitrarily high orders~\cite{Shchukin,Miranowicz}. 
Recently, on this basis \emph{genuine} non-Gaussian entanglement being invisible in second moments of the field quadratures, could be experimentally demonstrated~\cite{Walborn}. 

PT entanglement tests are necessary and sufficient for bipartite entanglement in very special cases only: for Hilbert spaces of dimension $2 \times 2$ and $2 \times 3$ and for Gaussian CV quantum states, cf. e.g.~\cite{Horodecki-09,Guehne-09}. Entanglement of general states 
can be verified by involved generalizations of the PT tests~\cite{Doherty},
or, based on entanglement witnesses\cite{Horodecki-96}, by optimized entanglement conditions~\cite{witness-sperling}. However, it has been proven that for any entangled state there exist subspaces of finite dimension in which the entanglement already exists~\cite{sperling-finite}. A systematic search for these subspaces provides the ability to uncover the structure of the entanglement. In particular, the identification of these entangled subsystems  may be helpful for designing protocols for the entanglement distillation, which extracts, in the ideal case, a maximally entangled state from the given mixed state, providing advantages in many quantum information tasks \cite{Horodecki-09}. 

In the present article we study the entanglement of a CV quantum state, experimentally created by mixing a fully phase-randomized squeezed vacuum state by a balanced beam splitter with vacuum noise. We analyze the entanglement structure of this decohered nonclassical state by PT tests on two-qubit subsystems. In this way, we extend the completeness of the Peres criterion for bipartite qubit systems to the detection of entanglement in arbitrary continuous-variable systems, and establish a link between entanglement in discrete variable and continuous variable systems. Several entangled qubits within the density matrix are identified with an overwhelming statistical significance. Surprisingly, the entanglement is only found between qubits of diverse nature. A closer inspection shows that the entanglement of these qubits can be distilled by the protocol of Bennett \emph{et al.} (BBPSSW)~\cite{bennett-dest}. 

The paper is structured as follows: In Section~\ref{sec:generate:state}, we present the experimental setup and describe the state whose entanglement is analyzed. In Section~\ref{sec:Gaussian:criteria}, we examine the detection of entanglement with Gaussian criteria, showing that they are not applicable for our states. The search for entangled qubit subsystems is given in Section~\ref{sec:entangled:qubits}, and its use for choosing a suitable distillation protocol is discussed. A summary and some conclusions are given in Sec.~\ref{concl}.

\section{Generation of a non-Gaussian entangled state.}\label{sec:generate:state}

\begin{figure}[h]
\includegraphics[width=\linewidth]{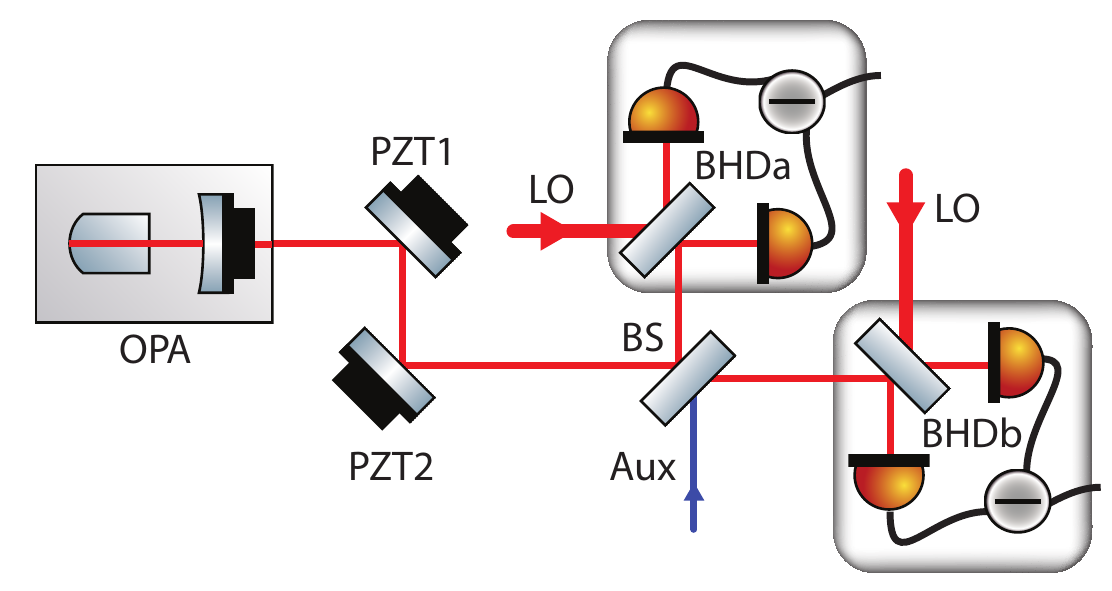}
\caption{Simplified sketch of the experimental setup. LO: local oscillator, OPA: optical parametric amplifier (squeezed light source), BHD: balanced homodyne detector, PZT: piezo-electrically actuated mirror applying the phase noise. Aux: Frequency shifted weak field providing a read-out for the difference of the detection phases of BHDa and BHDb.  }
\label{fig:setup}
\end{figure}

We start with a Gaussian squeezed-vacuum state being fully characterized by its variances $V_x$ and $V_p$ of the quadratures $x$ and $p$, respectively, together with its orientation angle $\varphi$ in phase space. To obtain a phase-independent non-Gaussian state, the orientation is uniformly distributed over a $2\pi$~interval. This state is not squeezed anymore, but still exhibits significantly nonclassical properties~\cite{PRA79-022122}, which are necessary for the generation of entanglement. In the second step, we send the phase-randomized state to a balanced  beam splitter where it is mixed with a vacuum mode. For a sketch of the experimental setup, see Fig.~\ref{fig:setup}. The two output modes, which form the entangled state examined in this article, are measured by joint balanced homodyne detectors. This scenario is equivalent to propagate an entangled two-mode squeezed vacuum through a medium preserving the phase difference but destroying the absolute phase, similar to the experiment in Ref.~\cite{zeilinger}. 

The squeezed mode was generated by a degenerate optical parametric
amplifier (OPA). The OPA consisted of a type-I non-critically phase
matched second order nonlinear crystal (7\% Mg:LiNbO$_3$) inside a
standing wave optical resonator with a line width of 25$\,$MHz. The OPA process was continuously pumped with second harmonic light at 532$\,$nm. The resonator tuning was controlled via the common Pound-Drever-Hall method using phase modulated fundamental light at 1064$\,$nm. For technical reason the pump phase was controlled such that the fundamental control field was deamplified, i.e. the OPA generated amplitude quadrature squeezing with respect to this field. With this setup we directly measured a
{squeezing} variance of -4.8$\,$dB and an {anti-squeezing}
variance of +9.0$\,$dB, both with respect to the unity vacuum noise variance. 

The squeezed field propagated in free space from the OPA passing high-reflection mirrors. In order to apply the phase diffusion two of these mirrors were moved by piezo-electric transducers (PZT) driven by a quasi-random voltage. The driving signal was generated by a high quality PC sound card connected to appropriate filters and amplifiers. The sound card played back a previously generated sound file which was carefully designed to yield the desired phase distribution.

Balanced homodyne detection (BHD) was used to measure the quadrature
amplitudes of the two-mode state under consideration. The visibility with the spatially filtered local oscillators were in the range of 98.5\% to 99\%. Regarding the detection phases the only meaningful figure is the difference between the detection phases of BHDa and BHDb because the state before the beamsplitter (BS) is phase randomized. The control of this phase difference was achieved by injecting an auxiliary field into the open port of the beamsplitter, which was frequency shifted with respect to the fundamental (LO) frequency. The demodulation of the BHD signal at the beat frequency provided an error signal for locking the detected quadrature angle to the phase of the auxiliary field with an offset given by the electronic demodulation phase. 
 The signals
of the two individual BHD-photodetectors were electronically mixed down
at 7$\,$MHz and low pass filtered with a bandwidth of 400$\,$kHz to
address a modulation mode of the light showing good squeezing and a high dark noise
clearance of the order of 20$\,$dB. The resulting signals were fed into
a PC based data acquisition system and sampled with one million samples
per second and 14 bit resolution.
 For a more detailed description of
the main parts of the setup we refer to~\cite{HageSQZpuri,FranzenSQZpuri}.

We obtained a set of $10^6$ quadrature pairs per detection phase configuration, for $10$ equally spaced phases per output mode $A$ and $B$. From these measurements, we estimated the density matrix elements of the state and their full covariance matrix via appropriate pattern functions, see Appendix~\ref{app:pattern} for details. 
Due to the central limit theorem, this yields the full statistical information about the experimental result.

\begin{figure}[h]
 	\includegraphics[width=\columnwidth]{{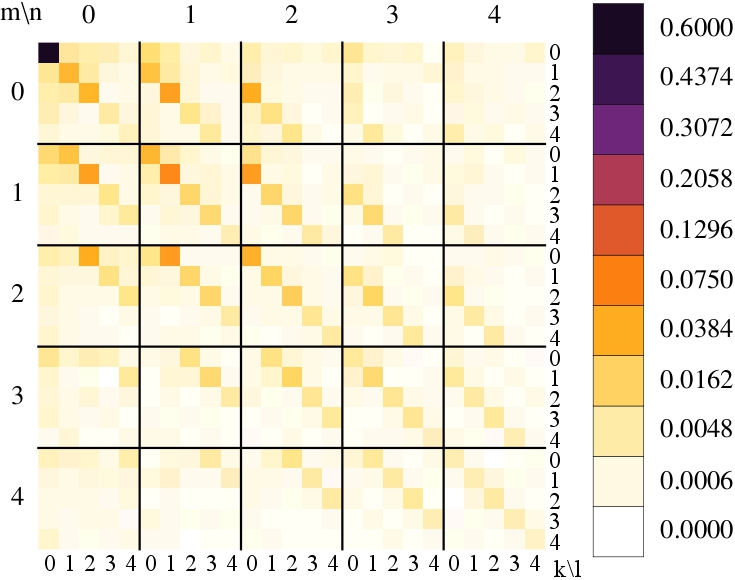}}
	\caption{Reconstructed two-mode density matrix of the examined entangled state in Fock basis. A block $mn$ provides the absolute values of the density matrix elements $\rho_{kl,mn}$ where $k, l$ refers to the photon numbers at BHDa, and $m, n$ to those at BHDb.}
	\label{fig:rho}
\end{figure}

The reconstructed density matrix of the measured state is shown in Fig.~\ref{fig:rho} for the first $5^4$ density matrix elements in the Fock basis. Our reconstruction is restricted to the $7$ lowest Fock states for each mode, thus the full bipartite density matrix has a rank of $7\times 7 = 49$. The trace of the reconstructed state equals to $0.9836\pm0.0002$, such that the main information about the state is covered. The standard deviation of the single matrix entries is bounded by $0.0003$.
It is noteworthy that a higher-dimensional state reconstruction makes no sense for the entanglement test. Already a reconstructed matrix of rank $8\times 8$ would consist of several entries that are no longer significant relative to the statistical error, as it is obtained for our sample.
The result of the state reconstruction shows that all significant matrix elements are located on certain lines. Only matrix elements with $k+m = l+n$ contribute, having their origin in the action of the beam splitter on the Fock state $\ket{k+m}$ in the input state. The phase diffusion eliminates all coherences between Fock states in the input state.

\section{Gaussian entanglement criteria.}\label{sec:Gaussian:criteria}
Entanglement of Gaussian states can be completely characterized by Simon's inequality~\cite{Simon}. Here we extend the notion of {\em Gaussian entanglement} to all quantum states for which the entanglement can be identified by the condition of Simon. This can be also the case for non-Gaussian states. Hence entangled states which cannot be identified by this condition will be denoted as \emph{genuinely} non-Gaussian ones.

In the following we prove that Simon's condition for entanglement is equivalent to squeezing of the input state. 
Let us start from a quadrature covariance matrix of a bipartite state in block form
\begin{equation}
  	V = \left(\begin{array}{c c} A & C \\ C^T & B\end{array}\right),
 \end{equation}
with $A,B$ being the covariance matrices of the subsystems, and define $J = \left(\begin{array}{c c} 0 & 1 \\ -1 & 0 \end{array}\right)$. The state itself does not have to be a Gaussian one, but we only examine its covariance matrix. A state is entangled if the following inequality is violated:
\begin{eqnarray}
 \det A \det B + &(\tfrac{1}{4} - |\det C|)^2 - {\rm Tr}{A J C J  B J C^T J} \nonumber\\
		&\geq \tfrac{1}{4}\left(\det A +\det B\right).
\end{eqnarray}
Let us assume that the covariance matrix of the initial state for the quadratures $\hat x_{\rm in},\hat p_{\rm in}$ is given by
\begin{equation}
 	C_{\rm in} = \left(\begin{array}{c c} V_x & C_{xp}\\ C_{xp} & V_p \end{array}\right).
\end{equation}
A beam splitter recombines the quadratures of this input field with the quadratures of vacuum to the quadratures of the two-mode output field $\hat x_3 = t \hat x_{\rm in} + r \hat x_{\rm vac}, \hat x_4 = -r \hat x_{\rm in} + t \hat x_{\rm vac}$, where the field transmissivity $t$ and reflectivity $r$ satisfy $|t|^2+|r|^2 =1$. Simon's criterion for the resulting covariance matrix leads to the following inequality for entanglement:
\begin{equation}
 	t^2 (1-t^2)\left[ (V_p-\tfrac{1}{2}) (V_x-\tfrac{1}{2})-C_{xp}^2\right] < 0. \label{eq:simon} 
\end{equation}
Since $0\leq t^2 \leq 1$, the term in square brackets has to be negative. In the following we show that this condition is equivalent to squeezing of the input state. The eigenvalues of the covariance matrix of the latter are the roots of the characteristic polynomial,
\begin{equation}
 	p(\lambda) = (V_x-\lambda)(V_p-\lambda) - C_{xp}^2.
\end{equation}
The two roots $\lambda_{1,2}$ are the minimum and maximum quadrature variance of the state. A state is squeezed if for one of these roots we find $\lambda_1 < \tfrac{1}{2}$, while for the other we have $\lambda_2 > \tfrac{1}{2}$. This is the case if and only if $p(\lambda)$ is negative between both roots, i.e.~$p(\tfrac{1}{2}) < 0$. This leads to
\begin{equation}
 	p(\tfrac{1}{2}) = (V_x-\tfrac{1}{2})(V_p-\tfrac{1}{2}) - C_{xp}^2 < 0,
\end{equation}
which is equivalent to Simon's condition for entanglement. Therefore, a beam splitter creates Gaussian entanglement if and only if the input state is squeezed. In this sense, one may state that Gaussian entanglement of the split state has its origin in Gaussian \emph{nonclassicality} of the input state.  We emphasize that this fact is not restricted to Gaussian states, but holds for arbitrary quantum states.

One can proceed with the analysis of higher-order moments. In \cite{Walborn}, a criterion containing fourth-order moments was sufficient to demonstrate genuine non-Gaussian entanglement. For the studied states the Simon test, based on second order-moments, failed. In our case, one can show that one has to go to sixth-order moments for this purpose, which is a cumbersome task. Therefore, we propose an alternative way to verify non-Gaussian entanglement, which does not only verify the entanglement. In addition it  provides useful insight into the entanglement structure of the state.

\section{Entangled qubits.}\label{sec:entangled:qubits}
Recently it has been been proven that any entangled quantum state must also be entangled in a finite dimensional Hilbert space~\cite{sperling-finite}. Therefore, it is sufficient to detect entanglement in a submatrix of the full density matrix. The Peres criterion states that for any separable state the partial transpose of the density matrix, defined by its matrix elements
\begin{equation}
 	\bra{k,m}\hat\rho^{\rm PT}\ket{l,n} = \bra{k,n}\hat \rho\ket{l,m},
\end{equation}
is positive semidefinite~\cite{Peres}. Hence, we may start to project the state onto an arbitrary two-qubit subsystem and calculate the eigenvalues of the partially transposed density matrix. If at least one of the latter is negative, the state has a negative partial transposition and entanglement has been verified by the Peres-criterion. This test can be implemented efficiently, since a bipartite qubit subsystem can be described by a $4\times4$ matrix, which has only four eigenvalues. The statistical uncertainty $\sigma(\lambda_{\rm min})$ of the eigenvalue $\lambda_{\rm min}$ can be obtained by a Monte-Carlo simulation: We draw 
random matrices, whose entries are chosen by a Gaussian distribution around the reconstructed state, with the covariance matrix as determined from the experimental data. With these simulated density matrices, which are consistent with the reconstructed state within its statistical uncertainties, we calculate the set of least eigenvalues. 
From the statistics of the results we can estimate the standard deviation of our experimental entanglement test. We note that whenever such a test fails for two-qubit subsystems it may be successful in subsystems of higher dimensions.

\begin{figure}
 \includegraphics[width=\columnwidth]{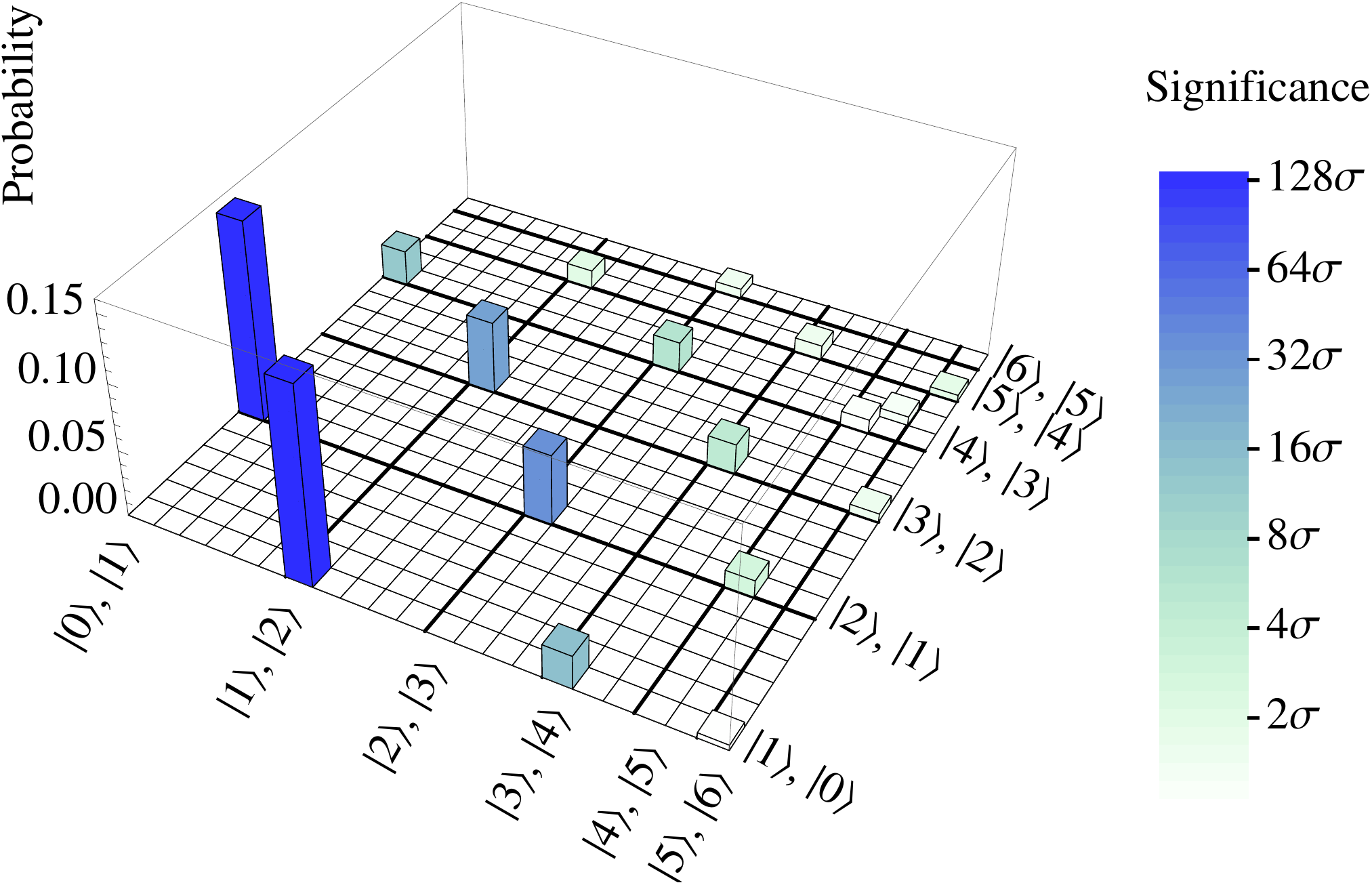} 
 \caption{Probabilities for the occurrence of entangled two-qubit subspaces. The statistical significance of the successful entanglement tests is color-coded. It represents the smallest (negative) eigenvalue of the partially transposed state in units of the corresponding standard deviation $\sigma$.}
	\label{fig:qubits}
\end{figure}

This test has been performed for all possible qubit subsystems. In Fig.~\ref{fig:qubits}, all two-qubit subsystems with negative partial transpose are illustrated by the probability of their occurrence in the state and by the statistical significance, $\tfrac{\lambda_{\rm min}}{\sigma(\lambda_{\rm min})}$, of the smallest eigenvalue. For each mode, the qubit subsystems are ordered from $\ket{0},\ket{1}$ to $\ket{0},\ket{6}$, continuing with $\ket{1},\ket{2}$ to $\ket{1},\ket{6}$ and so on. Each of the two axes running over subsystem labels refer to one of the two modes. First, it is obvious that there is no symmetric subsystem which displays the entanglement. That is, if both modes are projected onto the same subsystem, no entanglement is indicated. However, projecting on different subsystems, such as the ones composed of the states $\ket{0}, \ket{1}$ in one mode and $\ket{1},\ket{2}$ in the other, gives states whose entanglement can be verified with significance up to $128$ standard deviations. In total we find 10 asymmetric pairs of qubits whose entanglement has a statistical significance of more than two standard deviations. The asymmetry is remarkable, since the state itself is symmetric with respect to both modes. The knowledge of the structure of the entanglement of a given mixed quantum state, in particular the identification of the localization of entangled qubits within a complex CV state, is essential for applications.

Since we find entangled two-qubit subsystems in the state under study, our results clearly show that entanglement distillation is possible. Now the question for a suitable protocol arises. One scheme, which works for any two-qubit state, has been given in~\cite{horodecki-dest}. 
Our method as presented here identifies appropriate two-qubit Hilbert subspaces. Any subspace in which entanglement is found, may be chosen for distillation. The higher the probability of occurrence of the subsystem in the full state, the more efficient is the distillation.

\begin{figure}
 	\includegraphics[width=\columnwidth]{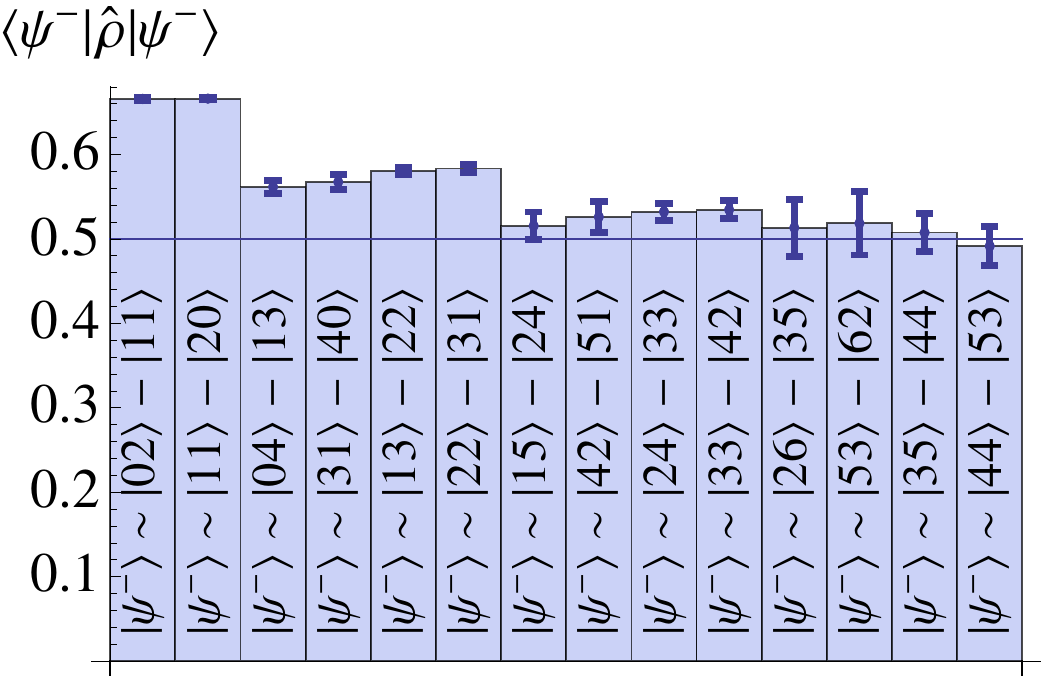}
	\caption{Probability of finding a singlet state $\ket{\psi^-}$ in a specific qubit subsystem.}
	\label{fig:fidelity}
\end{figure}

For some entangled qubits, one can directly apply the BBPSSW protocol~\cite{bennett-dest}.
Let us have a look at the fidelity of a singlet state with respect to the experimentally generated one within the chosen two-qubit subspace.  A singlet in this two-qubit space is defined as $\ket{\psi^-} = \frac{1}{\sqrt{2}}\left(\ket{i_A,j_B}-\ket{j_A,i_B}\right)$, where $\ket{i_A},\ket{j_A}$ and $\ket{i_B},\ket{j_B}$ are Fock states forming a basis for the qubit systems in the modes $A$ and $B$, respectively. If the fidelity between the state and the singlet, 
\begin{equation}
 	F =\bra{\psi^-} \hat \rho\ket{\psi^-},
\end{equation}
is greater than $\frac{1}{2}$, the BBPSSW protocol can be directly applied to distill the state. In Fig.~\ref{fig:fidelity}, the fidelities for the subsystems of interest are shown. Obviously, several subspaces have a fidelity which is significantly larger than $\frac{1}{2}$ being suitable for the application of the BBPSSW protocol.

\section{Summary and Conclusions}
\label{concl}
We have demonstrated an efficient method to examine the entanglement structure of  CV quantum states using the example of a mixed, genuine non-Gaussian entangled state whose entanglement is invisible for Gaussian entanglement criteria. We have analyzed the two-qubit subsystems and found negative partial transpositions with large statistical significances of up to 128 standard deviations. Our method proves that the CV state under consideration is entangled and distillable. 

A remarkable result of our analysis is that the decohered state under consideration \emph{only} shows entanglement in two-qubit subsystems composed of \emph{diverse} qubits, for instance  $\ket{0},\ket{1}$ and  $\ket{2},\ket{1}$. Such an insight into the entanglement structure might be used to analyze the decoherence process that turned an initially pure entangled state into a mixed entangled one. 
Moreover, the knowledge of the entanglement structure is important for the development of proper filtering methods for the aim of entanglement distillation.

\section*{Acknowledgments}
The authors are grateful to J.~Sperling, J.~DiGuglielmo and A.~Samblowski for fruitful discussions and their assistance. This work was supported by the Deutsche Forschungsgemeinschaft through SFB 652 and the Centre for Quantum Engineering and Space-Time Research, QUEST.

\appendix

\section{Reconstruction of the density matrix.}\label{app:pattern}

We obtained a set of $N=10^6$ quadrature pairs $\{x_j(\phi_{j_A}),y_j(\phi_{j_B})\}_{j=1}^{N}$ data points per phase configuration, for $N_\phi = 10$ equally spaced phases $\phi_{j_{A,B}} = \tfrac{j_{1,2}\pi}{10}$ per output mode $A,B$. From these measurements, we estimated the density matrix of the state via appropriate pattern functions,
\begin{eqnarray}
 	F^{(r)}_{kl,mn}(j,j_A,j_B) &=& {\rm Re}\big\{e^{i(k-l)\phi_{j_A}}f_{kl}(x_j(\phi_{j_A}))\nonumber\\
				&&\times e^{i(m-n)\phi_{j_B}}f_{mn}(y_j(\phi_{j_B}))\big\},\\
 	F^{(i)}_{kl,mn}(j,j_A,j_B) &=& {\rm Im}\big\{e^{i(k-l)\phi_{j_A}}f_{kl}(x_j(\phi_{j_A}))\nonumber\\
				&&\times e^{i(m-n)\phi_{j_B}}f_{mn}(y_j(\phi_{j_B}))\big\},
\end{eqnarray}
where $f_{kl}(x)$ has been taken from~\cite{PatternFunction}. The superscripts $(r)$, $(i)$ indicate the real and imaginary parts of the pattern functions. Then the real and imaginary part of the density matrix elements $\rho_{kl,mn} = \rho^{(r)}_{kl,mn} + i \rho^{(i)}_{kl,mn}$ are estimated as the empirical mean of $F_{kl,mn}(j,j_A,j_B)$:
\begin{equation}
 	\tilde\rho^{(r,i)}_{kl,mn} = \left< F^{(r,i)}_{mn,kl}(j,j_A,j_B)\right>,\label{eq:mean}
\end{equation}
where the brackets are symbols for 
\begin{equation}
	\left<F(j,j_A,j_B)\right> = \tfrac{1}{N_\phi^2 N}\sum_{j_A = 1}^{N_\phi}\sum_{j_B = 1}^{N_\phi} \sum_{j =1}^{N} F(j,j_A,j_B).
\end{equation}
Of course, all diagonal elements of the density matrix are real: $\tilde \rho_{kk,mm}^{(i)} = 0$. Furthermore, the covariance matrix of all entries can be estimated in the standard way as
\begin{widetext}
\begin{equation}
 {\rm Cov}\left(\tilde\rho_{kl,mn}^{(r,i)}\tilde\rho_{k'l',m'n'}^{(r,i)}\right) = 
	\frac{1}{N_\phi^2 N}\left[\left< F^{(r,i)}_{mn,kl}(j,j_A,j_B)F^{(r,i)}_{m'n',k'l'}(j,j_A,j_B)\right> - \tilde\rho_{kl,mn}^{(r,i)}\tilde\rho_{k'l',m'n'}^{(r,i)}\right].\label{eq:cov}
\end{equation}
\end{widetext}
Since the quadrature pairs are identically and independently distributed for any fixed pair of phases $\phi_{j_A},\phi_{j_B}$, the sums over the pattern functions are Gaussian distributed due to the central limit theorem. Therefore, Eqs~(\ref{eq:mean}) and (\ref{eq:cov}) 
provide the full statistical information about the sampled density matrix.


\begin{thebibliography}{52}
\bibitem{EPR} A.~Einstein, B.~Podolsky, and N.~Rosen, Phys. Rev. {\bf 47}, 777 (1935).
\bibitem{Naturw23-807} E.~Schr\"odinger, Naturwiss. {\bf 23}, 807 (1935).
\bibitem{Horodecki-09}  R.~Horodecki, P.~Horodecki, M.~Horodecki, and K.~Horodecki,  Rev. Mod. Phys. 81, 865–942 (2009).
\bibitem{Guehne-09} O.~G\"uhne and G.~T\'{o}th, Physics Reports {\bf 474}, 1 (2009).
\bibitem{Vedral} M. Arndt, T. Juffmann, and V. Vedral, HFSP Journal {\bf 3}, 386 (2009).
\bibitem{cvqubit} J.~S.~Neergaard-Nielsen, M.~Takeuchi, K.~Wakui, H.~Takahashi, K.~Hayasaka, M.~Takeoka, and M.~Sasaki, Phys. Rev. Lett. {\bf 105}, 053602 (2010).
\bibitem{Werner} R. F. Werner, Phys. Rev. A {\bf 40}, 4277 (1989).
\bibitem{Peres} A.~Peres, Phys. Rev. Lett. {\bf 77}, 1413 (1996).
\bibitem{Duan} L.M.~Duan, G.~Giedke, J.I.~Cirac, and P.~Zoller, Phys. Rev. Lett. {\bf 84}, 2722 (2000).
\bibitem{Simon} R.~Simon, Phys.~Rev.~Lett. {\bf 84}, 2726 (2000).

\bibitem{Shchukin} E.~Shchukin and W.~Vogel, Phys. Rev. Lett. {\bf 95}, 230502 (2005)
\bibitem{Miranowicz} A.~Miranowicz, M.~Piani, P.~Horodecki, and R.~Horodecki, Phys. Rev. A {\bf 80} 052303 (2009).
\bibitem{Walborn} R.~M.~Gomes, A.~Salles, F.~Toscano, P.~H.~Souto Ribeiro, and S.~P.~Walborn, Proc. Natl. Acad. Sci. U.S.A. 106, 21517 (2009).
\bibitem{Doherty} A.~C.~Doherty, P.~A.~Parrilo, and F.~M.~Spedalieri, Phys. Rev. Lett. {\bf 88}, 187904 (2002).
\bibitem{Horodecki-96} M.~Horodecki, P~Horodecki, and R.~Horodecki, Phys. Lett. A {\bf 223}, 1 (1996).
\bibitem{witness-sperling} J.~Sperling and W.~Vogel, Phys. Rev. A {\bf 79},022318 (2009).
\bibitem{sperling-finite} J.~Sperling and W.~Vogel, Phys. Rev. A {\bf 79}, 052313 (2009).
\bibitem{bennett-dest} C.~H. Bennett, G. Brassard, S. Popescu, B. Schumacher, J. A. Smolin, and W.~K. Wootters, Phys. Rev. Lett. {\bf 76}, 722 (1996).
\bibitem{PRA79-022122} T.~Kiesel, W.~Vogel, B.~Hage, J.~DiGuglielmo, A.~Samblowski, and R.~Schnabel, Phys. Rev. A {\bf 79}, 022122 (2009).
\bibitem{zeilinger} A. Fedrizzi et al., Nature Phys. {\bf 5}, 389 (2009). 
\bibitem{FranzenSQZpuri} A.~Franzen,  B.~Hage, J.~DiGuglielmo, J.~Fiur\'{a}\v{s}ek, and R.~ Schnabel, Phys. Rev. Lett. {\bf 97}, 150505 (2006).
\bibitem{HageSQZpuri} B.~Hage, A.~Samblowski, J.~DiGuglielmo, A.~Franzen, J.~Fiur\'{a}\v{s}ek, and R.~Schnabel, Nature Physics {\bf 4}, 915 (2008).

\bibitem{horodecki-dest} M. Horodecki, P. Horodecki, and R. Horodecki, Phys. Rev. Lett. {\bf 78}, 574 (1997); {\bf 80}, 5239 (1998).
\bibitem{PatternFunction} U.~Leonhardt, H.~Paul, and G.~M.~D'Ariano, Phys. Rev. A {\bf 52}, 4899 (1995).
\end{thebibliography}
\end{document}